\documentclass[useAMS,usenatbib,onecolumn]{mn2e}
\usepackage{amsmath,bm}
\usepackage{dcolumn}
\usepackage[utf8]{inputenc}
\usepackage{graphics,graphicx}
\usepackage{float}
\usepackage{threeparttable}
\usepackage{url}

\usepackage{color}

\newcommand{\p}{^\prime}
\newcommand{\pp}{^{\prime\prime}}

\newcommand{\bra}[1]{\langle #1|}
\newcommand{\ket}[1]{|#1\rangle}

\title[ExoMol line lists -- XXI. SiH$_4$]{ExoMol line lists -- XXII. The rotation-vibration spectrum of silane up to 1200\,K}

\date{\today}

\author[A. Owens et al.]
{A. Owens$^{1}$\thanks{The corresponding author: alec.owens.13@ucl.ac.uk}\thanks{Present address: The Hamburg Center for Ultrafast Imaging, Universit\"{a}t Hamburg, Luruper Chaussee 149, 22761 Hamburg, Germany}, A. Yachmenev$^2$, W. Thiel$^3$, J. Tennyson$^1$\thanks{The corresponding author: j.tennyson@ucl.ac.uk} and  S. N. Yurchenko$^1$\thanks{The corresponding author: s.yurchenko@ucl.ac.uk}\\
$^1$ Department of Physics and Astronomy, University College London, Gower Street, WC1E 6BT London, United Kingdom\\
$^2$ Center for Free-Electron Laser Science (CFEL), Deutsches Elektronen-Synchrotron DESY, Notkestrasse 85, 22607 Hamburg, Germany\\
$^3$ Max-Planck-Institut f\"{u}r Kohlenforschung, Kaiser-Wilhelm-Platz 1, 45470 M\"{u}lheim an der Ruhr, Germany}

\date{Accepted XXXX. Received XXXX; in original form XXXX}

\pagerange{\pageref{firstpage}--\pageref{lastpage}} \pubyear{2017}

\begin{document}

\label{firstpage}

\maketitle

\begin{abstract}
A variationally computed $^{28}$SiH$_4$ rotation-vibration line list applicable for temperatures up to $T=1200\,$K is presented. The line list, called OY2T, considers transitions with rotational excitation up to $J=42$ in the wavenumber range $0$--$5000\,$cm$^{-1}$ (wavelengths $\lambda> 2\,\mu$m). Just under 62.7 billion transitions have been calculated between 6.1 million energy levels. Rovibrational calculations have utilized a new `spectroscopic' potential energy surface determined by empirical refinement to 1452 experimentally derived energy levels up to $J=6$, and a previously reported \textit{ab initio} dipole moment surface. The temperature-dependent partition function of silane, the OY2T line list format, and the temperature dependence of the OY2T line list are discussed. Comparisons with the PNNL spectral library and other experimental sources indicate that the OY2T line list is robust and able to accurately reproduce weaker intensity features. The full line list is available from the ExoMol database and the CDS database.
\end{abstract}


\section{Introduction}

 The possibility of silicon-based life existing elsewhere in the Universe is an idea very much rooted in the realms of science fiction. Yet the continued discovery of exoplanets and the exhaustive search for potential biosignature gases~\citep{Seager:2016} has renewed interest in the simplest silicon-hydrogen compound, silane (SiH$_4$). A large number of high-resolution studies on the infrared (IR) absorption spectrum of SiH$_4$ and its isotopomers have been reported since the first measurements by \citet{34StNixx.SiH4,35StNixx.SiH4}. Rotation-vibration transitions have been detected around the carbon star IRC +10216~\citep{84GoBexx.SiH4,93KeRixx.SiH4,00MoDaHa.SiH4}, and in the atmospheres of Jupiter~\citep{78TrLaFi.SiH4} and Saturn~\citep{80LaFiSm.SiH4}. Although deemed unlikely, silane has already been considered in the context of H$_2$-dominated atmospheres on rocky exoplanets~\citep{13aSeBaHu.CH3Cl}.

 Despite efforts to document the spectrum of SiH$_4$, no complete line list exists for this molecule. The PNNL spectral library~\citep{PNNL} contains IR absorption cross-sections at a resolution of around $0.06\,$cm$^{-1}$ for the $600$--$6500\,$cm$^{-1}$ range. Another useful resource is the Spherical Top Data System (STDS)~\citep{STDS:1998}, however, a significant portion of the measured transitions and intensities are from unpublished work which makes it difficult to assess the reliability of the data. Theoretical $^{28}$SiH$_4$ spectra are available from the TheoReTS database~\citep{TheoReTS:2016} for a temperature range of $70$--$300\,$K but the calculations that they are based on are again from unpublished work. A room temperature line list has been produced for the $750$--$1150\,$cm$^{-1}$ region from analysis of the $\nu_2$ and $\nu_4$ bands~\citep{17UlGrBe.SiH4}. Whilst absolute line intensities have been determined by \citet{15HeLoNa.SiH4} for a large number of P-branch transitions of the $\nu_3$ band up to the rotational quantum number $J=16$.

 Previously, we generated potential energy and dipole moment surfaces for silane using state-of-the-art \textit{ab initio} theory~\citep{15OwYuYa.SiH4}. Both surfaces were rigorously evaluated and showed good agreement with experimental results. Computed fundamental term values of $^{28}$SiH$_4$ using the CBS-F12$^{\,\mathrm{HL}}$ potential energy surface (PES) possessed a root-mean-square (rms) error of $0.63\,$cm$^{-1}$ compared to experiment. The dipole moment surface (DMS), although tending to marginally overestimate the strength of line intensities, reproduced band shape and structure well.

 Building on this work, we present a comprehensive rotation-vibration line list of $^{28}$SiH$_4$ suitable for elevated temperatures. The line list, called OY2T, has been computed variationally and utilizes a new `spectroscopic' PES which has been determined by rigorous empirical refinement. The OY2T line list has been produced for the ExoMol database~\citep{ExoMol2012,ExoMol2016}, which is providing important molecular spectroscopic data to help characterize exoplanet and other hot atmospheres. Examples of the application of these line lists include: the early detection of water in HD~189733b~\citep{jt400} and HD~209458b~\citep{jt488} using the BT2 line list~\citep{jt378}; the tentative identification of HCN in the atmosphere of super-Earth 55 Canri e~\citep{jt629} and TiO in the atmosphere of hot Jupiter WASP-76 b~\citep{jt699}; various studies using the 10to10 line list~\citep{14YuTexx.CH4} to detect methane in exoplanets~\citep{jt495,14YuTeBa.CH4,jt699}, the bright T4.5 brown dwarf 2MASS 0559-14~\citep{14YuTeBa.CH4}, and to make detailed line assignments in the near-infrared spectra of late T dwarfs~\citep{jt596} in conjunction with the BYTe ammonia line list~\citep{jt500}. Conversely, a line list for the diatomic NaH molecule~\citep{jt605} was able to rule out the tentative detection of this species in the atmosphere of a brown dwarf.

 Silane is the second five-atom molecule to be treated within the ExoMol framework~\citep{ExoSoft2016} after methane and the 10to10 line list~\citep{14YuTexx.CH4}, which demonstrated the need to consider a very large number of transitions to correctly model the opacity at elevated temperatures~\citep{14YuTeBa.CH4}. Since the 10to10 line list, several key developments have taken place in our nuclear motion code \textsc{trove}~\citep{TROVE2007,15YaYu.ADF} that significantly ease the computational burden of theoretical line list production. These include an automatic differentiation method to construct the rovibrational Hamiltonian, the implementation of curvilinear internal coordinates, and a novel vibrational basis set truncation approach which will be discussed later on.

  The paper is structured as follows: In Sec.~\ref{sec:methods} we detail the empirical refinement of the CBS-F12$^{\,\mathrm{HL}}$ PES, the DMS and intensity simulations, and the variational calculations. In Sec.~\ref{sec:results}, the OY2T line list is presented and we discuss the temperature-dependent partition function of SiH$_4$, the format of the OY2T line list, the temperature dependence of the OY2T line list, and comparisons with the PNNL spectral library and other experimental data. Concluding remarks are given in Sec.~\ref{sec:conc}.

\section{Methods}
\label{sec:methods}

\subsection{Potential energy surface refinement}

 The CBS-F12$^{\,\mathrm{HL}}$ PES~\citep{15OwYuYa.SiH4} was constructed from extensive, explicitly correlated coupled cluster calculations with extrapolation to the complete basis set limit, and incorporated additional higher-level energy corrections to account for core-valence electron correlation, higher-order coupled cluster terms beyond CCSD(T), and scalar relativistic effects. Although impressive in its accuracy, orders-of-magnitude improvements in predicted transition frequencies can be achieved by refining the PES to experiment. Furthermore, improved energy levels means better wavefunctions and more reliable intensities.

 The refinement was carried out using an efficient least-squares fitting procedure~\citep{YuBaTe11.NH3} implemented in \textsc{trove}. The procedure works by assuming the CBS-F12$^{\,\mathrm{HL}}$ PES, $V_{\rm CBS\mbox{-}F12^{HL}}$, is already a reasonable representation such that the effect of the refinement can be treated as a perturbation $\Delta V$, i.e.
\begin{equation}
V^{\prime}=V_{\rm CBS\mbox{-}F12^{HL}}+\Delta V ,
\end{equation}
where $V^{\prime}$ is the refined PES. The correction $\Delta V$ is expanded in terms of nine vibrational coordinates ${\bm\xi}=\lbrace\xi_1,\xi_2,\ldots,\xi_9\rbrace$ according to the formula,
\begin{equation}
\Delta V = \sum_{ijk\ldots} \Delta \mathrm{f}_{ijk\ldots}\left\lbrace\xi_1^i\xi_2^j\xi_3^k\ldots\right\rbrace^{A_1} ,
\end{equation}
where the coefficients $\Delta \mathrm{f}_{ijk\ldots}$ are corrections to the original PES expansion parameters $\mathrm{f}_{ijk\ldots}$. The expansion terms $\left\lbrace\xi_1^i\xi_2^j\xi_3^k\ldots\right\rbrace^{A_1}$ are symmetrized combinations of the vibrational coordinates ${\bm\xi}=\lbrace\xi_1,\xi_2,\ldots,\xi_{9}\rbrace$ and transform according to the $A_1$ irreducible representation of the $\bm{T}_{\mathrm{d}}\mathrm{(M)}$ molecular symmetry group~\citep{MolSym_BuJe98}. Further details on the vibrational coordinates and analytic representation of the CBS-F12$^{\,\mathrm{HL}}$ PES can be found in \citet{15OwYuYa.SiH4}.

  The new perturbed Hamiltonian, $H^{\prime}=H+\Delta V$, is then diagonalized using a basis set of eigenfunctions from the initial unperturbed Hamiltonian $H$ eigenvalue problem. Each iteration utilizes the previous ``unperturbed'' basis set in this way until a PES of desirable quality is reached. To ensure the consistency of the refined surface, the expansion parameters $\mathrm{f}_{ijk\ldots}$ are simultaneously fitted to both the experimental data and the original \textit{ab initio} dataset~\citep{03YuCaJe.PH3}. This stops any unrealistic distortion of the PES in regions not sampled by experiment. It also means each expansion parameter can be adjusted irrespective of the amount, or distribution, of experimental energies used in the refinement.

 For SiH$_4$, a total of 1452 experimental term values up to $J=6$ were used to refine the CBS-F12$^{\,\mathrm{HL}}$ PES. This included 53 vibrational $J=0$ band centres from \citet{STDS:1998,Chevalier:1988,91ZhQiMa.SiH4,95SuWaZh.SiH4,98LiWaCh.SiH4}, and 1399 $J>0$ energies from the STDS~\citep{STDS:1998}. Of the $106$ expansion parameters of the CBS-F12$^{\,\mathrm{HL}}$ PES, $104$ were varied and the results of the refinement are shown in Fig.~1, where we have plotted the fitting residuals $\Delta E({\rm obs-calc})=E_{\rm obs}-E_{\rm calc}$. Here, $E_{\rm obs}$ and $E_{\rm calc}$ are the observed and calculated energies, respectively. Overall, the experimental energy levels are reproduced with a rms error of $0.454\,$cm$^{-1}$. This error is slightly larger than what is achievable with refinement, however, this was intentional. The energy levels from the STDS are derived from an effective Hamiltonian model based on some unpublished work. It is therefore hard to assess the reliability of the data and how much it can be trusted in the refinement. Despite this, because the CBS-F12$^{\,\mathrm{HL}}$ PES is reliable we could depend on it more and ensure the refined PES did not deviate too far away from the original surface.

\begin{figure}
\centering
\label{fig:refine}
\includegraphics[width=0.47\textwidth]{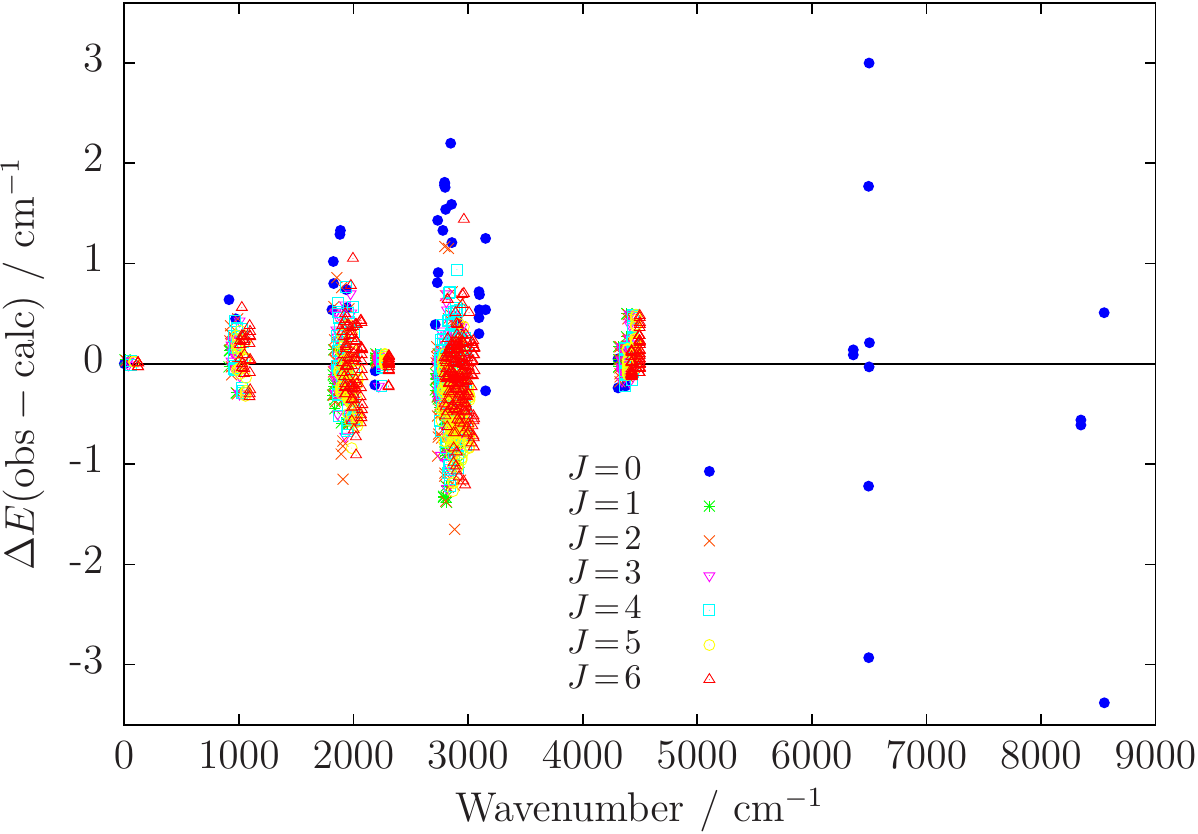}
\caption{Fitting residuals, $\Delta E({\rm obs-calc})=E_{\rm obs}-E_{\rm calc}$, of the 1452 energy levels used in the PES refinement.}
\end{figure}

 Pure rotational energies were given the largest weights in the refinement, while $J=0$ and rovibrational term values were weighted two orders of magnitude smaller with values between $1$--$10$ based on their energy. Since the weights are normalized in the fitting, relative weighting is more important than absolute values (see \citet{YuBaTe11.NH3} for further details). Although we have determined a new `spectroscopic' PES of silane, the accuracy of the refined PES is only guaranteed with the computational setup employed in this study. This is usually the case in theoretical line list production using programs that do not treat the kinetic energy operator exactly, and we therefore do not recommend this PES for use otherwise.

\subsection{Dipole moment surface and line intensities}

 The electric DMS utilized for the present study was computed at the CCSD(T)/aug-cc-pVTZ(+d for Si) level of theory. Further details on the analytic representation and performance of the DMS can be found in \citet{15OwYuYa.SiH4}, where it was shown that the DMS marginally overestimates the strength of line intensities. This discrepancy is very minor though as we will see in Sec.~\ref{sec:linelist}.

 To simulate absolute absorption intensities we used the expression,
\begin{equation}
\label{eq:abs_I}
I(f \leftarrow i) = \frac{A_{if}}{8\pi c}g_{\mathrm{ns}}(2 J_{f}+1)\frac{\exp\left(-E_{i}/kT\right)}{Q(T)\; \nu_{if}^{2}}\left[1-\exp\left(-\frac{hc\nu_{if}}{kT}\right)\right] ,
\end{equation}
where $A_{if}$ is the Einstein $A$ coefficient of a transition with wavenumber $\nu_{if}$ (in cm$^{-1}$) between an initial state with energy $E_i$, and a final state with rotational quantum number $J_f$. Here, $k$ is the Boltzmann constant, $T$ is the absolute temperature, $h$ is the Planck constant and $c$ is the speed of light. The nuclear spin statistical weights are $g_{\mathrm{ns}}=\lbrace 5,5,2,3,3\rbrace$ for states of symmetry $\lbrace A_1,A_2,E,F_1,F_2\rbrace$, respectively. The partition function for a given temperature $Q(T)$ will be discussed in Sec.~\ref{sec:pfn}. All transitions obey the symmetry selection rules
\begin{equation}
A_1 \leftrightarrow A_2,\; E \leftrightarrow E,\; F_1 \leftrightarrow F_2 ,
\end{equation}
and the standard rotational selection rules,
\begin{equation}
J\p-J\pp=0,\pm 1,\; J\p+J\pp \ne 0 ,
\end{equation}
where $\p$ and $\pp$ denote the upper and lower state, respectively. Note that the \textsc{ExoCross} code (Yurchenko, Al-Refaie \& Tennyson (in preparation); available at \url{https://doi.org/10.5281/zenodo.400748}) was employed for all spectral simulations.

\subsection{Variational calculations}

 The computer program \textsc{trove} was used for all rovibrational calculations. Since the methodology of \textsc{trove} is well documented~\citep{TROVE2007,09YuBaYa.NH3,15YaYu.ADF,Symmetry:2017} and we have previously reported calculations on SiH$_4$~\citep{15OwYuYa.SiH4}, we summarise only the key aspects relevant for this work.

 The rovibrational Hamiltonian was constructed numerically using an automatic differentiation method~\citep{15YaYu.ADF}. The Hamiltonian was represented as a power series expansion around the equilibrium geometry in terms of nine, curvilinear internal coordinates, with the kinetic and potential energy operators truncated at 6th order. Atomic mass values were used throughout. A multi-step contraction scheme was used to build the vibrational basis set, the size of which was controlled by the polyad number
\begin{equation}\label{eq:polyad_sih4}
P = 2(n_1+n_2+n_3+n_4)+n_5+n_6+n_7+n_8+n_9 \leq P_{\mathrm{max}} .
\end{equation}
The quantum numbers $n_k$ for $k=1,\ldots,9$ are related to primitive basis functions $\phi_{n_k}$, which are obtained by solving a one-dimensional Schr\"{o}dinger equation for each $k$th vibrational mode with the Numerov-Cooley method~\citep{Numerov1924,Cooley1961}. Multiplication with symmetrized rigid-rotor eigenfunctions $\ket{J,\Gamma_{\mathrm{rot}},n}$ produces the final basis set for $J>0$ calculations. Here, the label $\Gamma_{\mathrm{rot}}$ is the rotational symmetry and $n$ is a multiplicity index used to count states within a given $J$ (see \citet{06BoReLo.CH4}).

 In calculations we set $P_{\mathrm{max}}=12$, which resulted in 19,237 vibrational basis functions corresponding to energies up to $h c \cdot 15$,000$\,$cm$^{-1}$. Describing high rotational excitations with such a large basis set can quickly become computationally intractable (rovibrational matrices scale linearly with $J$). It was therefore necessary to reduce the number of basis functions. To do this we employed a new, basis set truncation approach (Yurchenko et al. (in preparation)) based on vibrational transition moments, which are relatively inexpensive to compute in \textsc{trove}. The vibrational transition moment between two states is defined as,
\begin{equation}
\label{eq:TM}
\mu_{if} = \sqrt{\sum_{\alpha=x,y,z}{\lvert\bra{\Phi^{(f)}_{\mathrm{vib}}}\bar{\mu}_{\alpha}\ket{\Phi^{(i)}_{\mathrm{vib}}}\rvert}^2} ,
\end{equation}
where $\ket{\Phi^{(i)}_{\mathrm{vib}}}$ and $\ket{\Phi^{(f)}_{\mathrm{vib}}}$ are the initial and final state vibrational eigenfunctions, respectively, and $\bar{\mu}_{\alpha}$ is the electronically averaged dipole moment function along the molecule-fixed axis $\alpha=x,y,z$.

 After computing $J=0$ energies, all possible transition moments were calculated for a lower energy threshold of $h c \cdot 8000\,$cm$^{-1}$ (same as for the final OY2T line list). Once known, we estimated the vibrational band intensity at an elevated temperature (e.g. $1500\,$K) for every transition moment. Each vibrational energy level, and consequently each $J=0$ basis function, was assigned a band intensity value which was simply the largest computed for that state when acting as $\ket{\Phi^{(i)}_{\mathrm{vib}}}$ in Eq.~\eqref{eq:TM}. The $J=0$ basis set was then pruned by removing basis functions and energy levels above $h c \cdot 8000\,$cm$^{-1}$ with band intensity values smaller than $1\times 10^{-21}\,$cm/molecule. This is around three orders of magnitude smaller than the largest computed value. The band intensity threshold was chosen solely on a computational basis so that the rovibrational calculations were manageable with the resources available to us. The resulting pruned basis set contained 3817 functions (corresponding to energies up to $h c \cdot 15$,000$\,$cm$^{-1}$) and was multiplied in the usual manner with symmetrized rigid-rotor functions for $J>0$ calculations.

 An advantage of this truncation approach is that the accuracy of the vibrational energy levels and respective wavefunctions computed using a larger value of $P_{\mathrm{max}}$ is retained. Naturally we will lose some information on weaker lines involving states above $h c \cdot 8000\,$cm$^{-1}$ and it is hard to quantify the effect without more rigorous calculations. Removing functions from the basis set will also influence predicted rovibrational energies but the errors introduced, and the basis set convergence error, can to some extent be compensated for by refining the PES with the pruned basis set, which is what we have done. The resulting `effective' PES is only reliable with the same computational setup but this is usually the case in theoretical line list production with \textsc{trove}.

 The OY2T line list was computed with a lower state energy threshold of $h c \cdot 8000{\,}$cm$^{-1}$ and considered transitions up to $J=42$ in the $0$--$5000{\,}$cm$^{-1}$ range. To improve the accuracy of the line list we employed an empirical basis set correction (EBSC)~\citep{09YuBaYa.NH3}, which is essentially a shift of the band centres to better match experiment. Recall that in the multi-step contraction scheme used by \textsc{trove}, the $J=0$ eigenfunctions form the final basis set and the vibrational part of the Hamiltonian is diagonal with respect to this basis set. It is possible then to replace the diagonal elements with accurate experimental values for the $J=0$ energies when constructing the Hamiltonian matrix. For the OY2T line list, only the four fundamentals, listed in Table~1, were replaced as these wavenumbers are well established.

\begin{table}
\centering
\tabcolsep=4pt
\caption{\label{tab:vibrations}Vibrational modes of SiH$_4$ and the observed band centres (in cm$^{-1}$) from \citet{STDS:1998}.}
\begin{tabular}{c c c l}
\hline\hline\\[-3.0mm]
Mode & Symmetry & Band Centre & Description \\[0.5mm]
\hline\\[-3.0mm]
$\nu_1$ & $A_1$ & 2186.87 & Symmetric stretching\\
$\nu_2$ & $E  $ &  \,\,\,970.93 & Antisymmetric bending\\
$\nu_3$ & $F_2$ & 2189.19 & Antisymmetric stretching\\
$\nu_4$ & $F_2$ &  \,\,\,913.47 & Antisymmetric bending\\[0.5mm]
\hline\hline
\end{tabular}
\end{table}

\section{Results}
\label{sec:results}

\subsection{Partition function of silane}
\label{sec:pfn}

 The temperature-dependent partition function $Q(T)$ is defined as,
\begin{equation}
\label{eq:pfn}
Q(T)=\sum_{i} g_i \exp\left(-\frac{E_i}{kT}\right) ,
\end{equation}
where $g_i=g_{\rm ns}(2J_i+1)$ is the degeneracy of a state $i$ with energy $E_i$ and rotational quantum number $J_i$. Summing over all computed rovibrational energy levels, in Fig.~2 we have plotted the convergence of $Q(T)$ as a function of $J$ for different temperatures. At $T=1200\,$K the partition function is converged to around $0.1\%$. Our calculated room temperature partition function $Q(296\,{\rm K})=1532.93$ is in excellent agreement with a value of $Q(296\,{\rm K})=1533.00$ from the TheoReTS database~\citep{TheoReTS:2016}. It also agrees well with an approximate estimate of $Q\approx Q_{\mathrm{rot}}\times Q_{\mathrm{vib}}=1527.36$ used in our previous study~\citep{15OwYuYa.SiH4}, where $Q_{\mathrm{rot}}$ and $Q_{\mathrm{vib}}$ are the rotational and vibrational partition function, respectively.

\begin{figure}
\centering
\label{fig:pfn}
\includegraphics[width=0.47\textwidth]{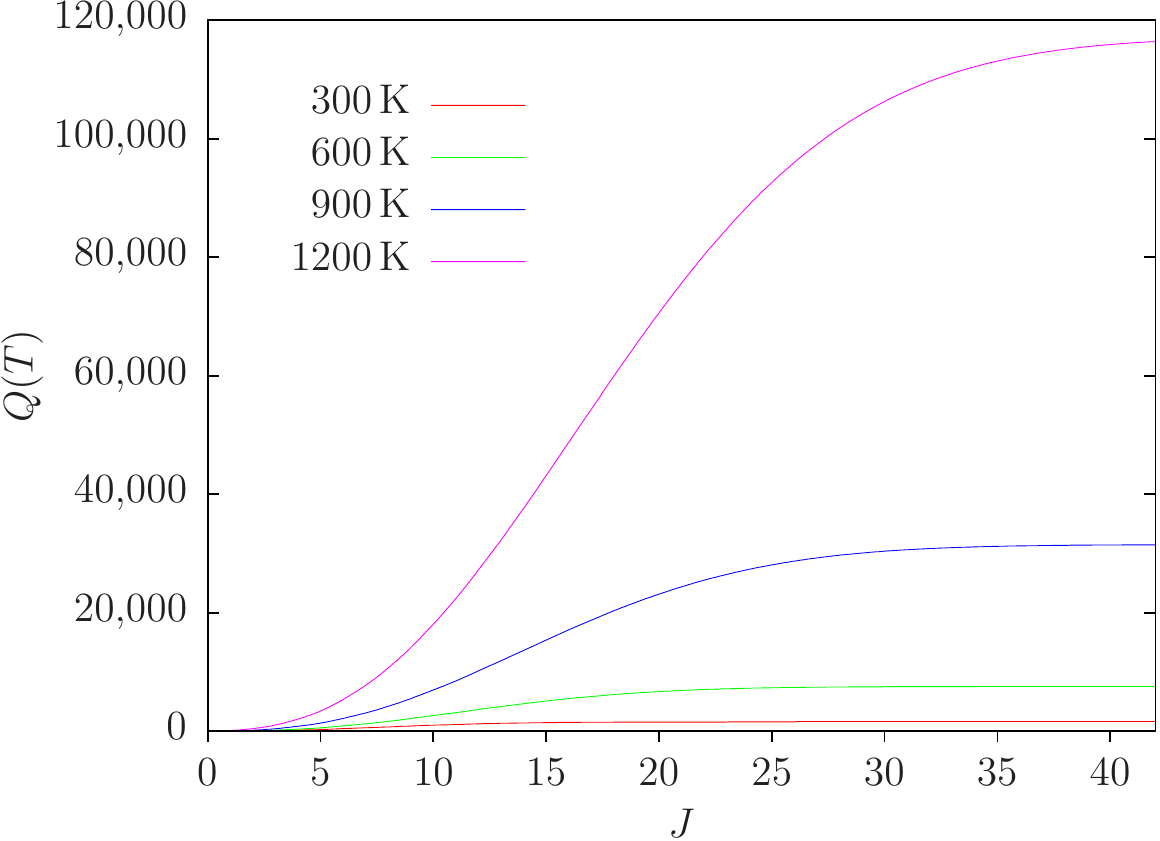}
\caption{Convergence of the partition function $Q(T)$ with respect to the rotational quantum number $J$ for different temperatures.}
\end{figure}

 Given that the OY2T line list has been computed with a lower energy threshold of $h c \cdot 8000{\,}$cm$^{-1}$, it is informative to  study a reduced partition function $Q_{\rm limit}$, which only considers energy levels up to $h c \cdot 8000{\,}$cm$^{-1}$ in the summation of Eq.~\eqref{eq:pfn}. In Fig.~3 we plot the ratio $Q_{\rm limit}/Q$ with respect to temperature, and this can provide a measure of completeness of the OY2T line list. At $T=1200\,$K, the ratio $Q_{\rm limit}/Q=0.94$ and we recommend this as a `soft' temperature limit to the OY2T line list. Using the line list above this temperature will result in the progressive loss of opacity, however, it is possible to estimate the missing contribution with the ratio $Q_{\rm limit}/Q$~\citep{Neale:1996}. Note that our full partition function evaluated on a $1\,$K grid from $70$--$1400\,$K is provided as supplementary material.

\begin{figure}
\centering
\label{fig:pfn_lim}
\includegraphics[width=0.47\textwidth]{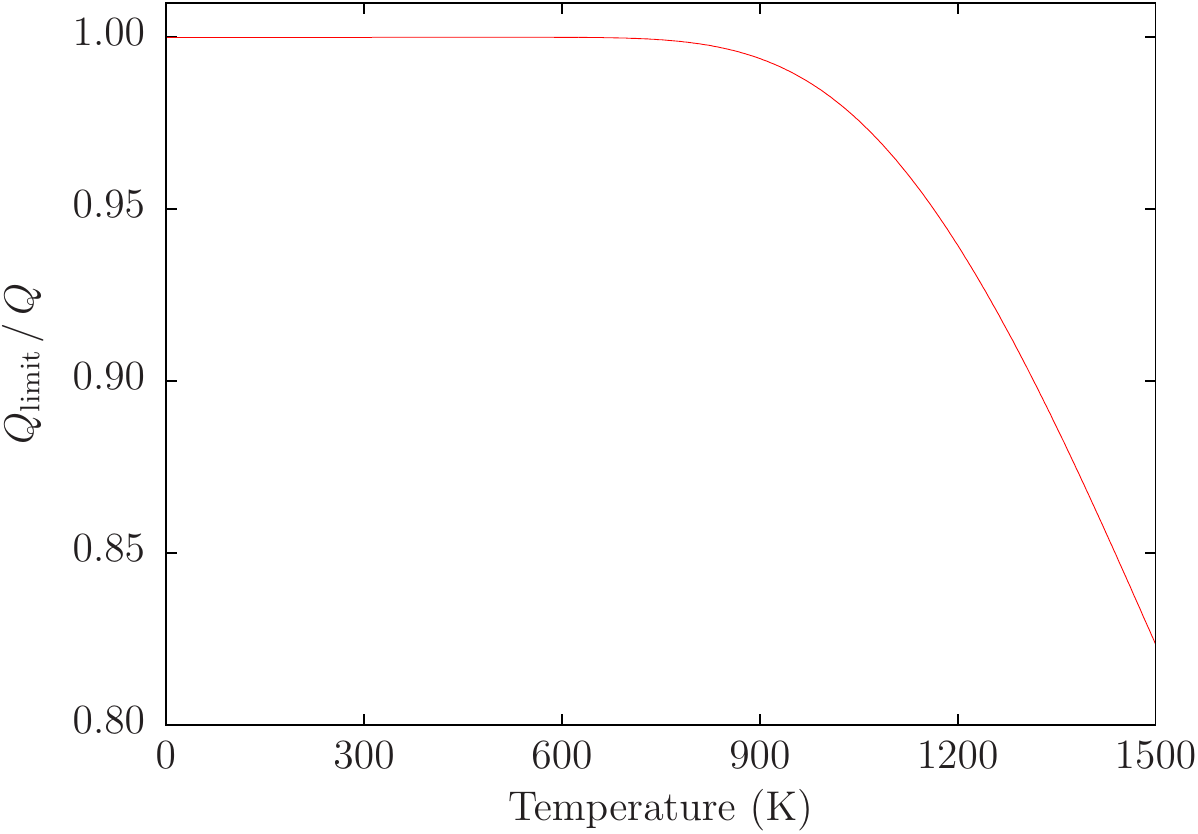}
\caption{Plot of the ratio $Q_{\rm limit}/Q$ as a function of temperature. This ratio provides a measure of completeness for the OY2T line list.}
\end{figure}

\subsection{OY2T line list format}

 A complete description of the ExoMol data structure along with examples was recently reported by \citet{ExoMol2016}. The \texttt{.states} file contains all computed rovibrational energies (in cm$^{-1}$). Each energy level possesses a unique state ID with symmetry and quantum number labelling as shown in Table~2. The \texttt{.trans} files, which are split into frequency windows so that they are easier to handle, contain all computed transitions with upper and lower state ID labels, and Einstein $A$ coefficients. An example from a \texttt{.trans} file for the OY2T line list is given in Table~3.

\begin{table*}
\caption{Extract from the \texttt{.states} file for the OY2T line list.}
\vspace*{-3mm}
\begin{threeparttable}
\label{tab:states}
\centering
\tabcolsep=5pt
\begin{tabular}{lrccccccccccccccccccc}
\hline\hline\\[-3mm]
        $N$  &  \multicolumn{1}{c}{$\tilde{E}$}   &  $g_{\rm tot}$  &  $J$ & $\Gamma_{\rm tot}$ & $n_1$ & $n_2$ & $n_3$ & $n_4$ & $n_5$ & $n_6$ & $n_7$ & $n_8$ & $n_9$ & $\Gamma_{\rm vib}$ & $J$ & $K$ & $\tau_{\rm rot}$ & $\Gamma_{\rm rot}$ & $ N_{J,\Gamma}$ & $|C_i^{2}|$ \\
\hline\\[-3mm]
 1&     0.000000&  5&  0&  1&  0&  0&  0&  0&  0&  0&  0&  0&  0&  1&  0&  0&  0&  1&  0&  0.99\\
 2&  1810.753814&  5&  0&  1&  0&  0&  0&  0&  0&  0&  2&  0&  0&  1&  0&  0&  0&  1&  0&  0.32\\
 3&  1936.931391&  5&  0&  1&  0&  0&  0&  0&  2&  0&  0&  0&  0&  1&  0&  0&  0&  1&  0&  0.49\\
 4&  2186.873254&  5&  0&  1&  0&  0&  1&  0&  0&  0&  0&  0&  0&  1&  0&  0&  0&  1&  0&  0.24\\
 5&  2730.476494&  5&  0&  1&  0&  0&  0&  0&  0&  0&  1&  1&  1&  1&  0&  0&  0&  1&  0&  0.91\\
 6&  2792.635361&  5&  0&  1&  0&  0&  0&  0&  1&  0&  2&  0&  0&  1&  0&  0&  0&  1&  0&  0.49\\
 7&  2914.995074&  5&  0&  1&  0&  0&  0&  0&  1&  2&  0&  0&  0&  1&  0&  0&  0&  1&  0&  0.73\\
 8&  3099.093127&  5&  0&  1&  0&  0&  1&  0&  0&  0&  0&  0&  1&  1&  0&  0&  0&  1&  0&  0.69\\
 9&  3599.518736&  5&  0&  1&  0&  0&  0&  0&  0&  0&  2&  0&  2&  1&  0&  0&  0&  1&  0&  0.20\\
10&  3651.898227&  5&  0&  1&  0&  0&  0&  0&  0&  0&  0&  4&  0&  1&  0&  0&  0&  1&  0&  0.21\\
\hline\hline
\end{tabular}
\begin{tablenotes}
\item $N$: State ID;
\item $\tilde{E}$: Term value (in cm$^{-1}$);
\item $g_{\rm tot}$: Total degeneracy;
\item $J$: Rotational quantum number;
\item $\Gamma_{\rm tot}$: Total symmetry in $\bm{T}_{\mathrm{d}}\mathrm{(M)}$ (1 is $A_1$, 2 is $A_2$, 3 is $E$, 4 is $F_1$, 5 is $F_2$);
\item $n_1$ -- $n_9$: \textsc{trove} vibrational quantum numbers;
\item $\Gamma_{\rm vib}$: Symmetry of the vibrational contribution in $\bm{T}_{\mathrm{d}}\mathrm{(M)}$;
\item $J$: Rotational quantum number (same as column 4);
\item $K$: Rotational quantum number, projection of $J$ onto molecule-fixed $z$-axis;
\item $\tau_{\rm rot}$: Rotational parity (0 or 1);
\item $\Gamma_{\rm rot}$: Symmetry of the rotational contribution in $\bm{T}_{\mathrm{d}}\mathrm{(M)}$;
\item $N_{J,\Gamma}$: State number in $J,\Gamma$ block;
\item $|C_i^{2}|$: Largest coefficient used in the assignment.
\end{tablenotes}
\end{threeparttable}
\end{table*}

\begin{table}
\centering
\tabcolsep=10pt
\caption{Extract from a \texttt{.trans} file for the OY2T line list.}
\label{tab:trans}
\begin{tabular}{rrr}
\hline\hline\\[-3mm]
       \multicolumn{1}{c}{$f$}  &  \multicolumn{1}{c}{$i$} & \multicolumn{1}{c}{$A_{if}$}\\
\hline\\[-3mm]
     1009016    &   887497 & 2.8705e-05 \\
     1012938    &   889497 & 1.6387e-05 \\
     1013144    &  1138036 & 5.4460e-02 \\
     1013796    &  1138422 & 1.0351e-01 \\
      101385    &    61971 & 5.4116e-04 \\
     1014053    &   890062 & 1.6747e-04 \\
     1014363    &  1010642 & 4.0593e-03 \\
     1016000    &  1139649 & 5.9798e-02 \\
     1017478    &   892208 & 1.0542e-06 \\
     1017606    &  1140576 & 5.3820e-03 \\
\hline\hline\\[-2mm]
\end{tabular}

\noindent
\footnotesize{
$f$: Upper  state ID; $i$:  Lower state ID; \\
$A_{if}$:  Einstein $A$ coefficient (in s$^{-1}$).}
\end{table}

\subsection{Validation of the OY2T line list}
\label{sec:linelist}

 The OY2T line list contains nearly 62.7 billion (62,690,449,078) transitions between 6.1 million (6,142,521) energy levels. The distribution of lines and energies is illustrated in Fig.~4, where we have plotted the total number computed for each value of $J$. The density of transitions is largest between $18\leq J \leq 26$ but this drops off relatively smoothly and by $J=42$ we have nearly calculated all possible transitions for our computational setup (e.g. lower energy threshold of $h c \cdot 8000{\,}$cm$^{-1}$, wavenumber range of $5000{\,}$cm$^{-1}$, pruned rovibrational basis set, etc.). The decrease in the number of energy levels after $J=34$ is a result of the upper energy threshold of $h c \cdot 13$,000{\,}cm$^{-1}$

\begin{figure}
\centering
\label{fig:distribution}
\includegraphics[width=0.47\textwidth]{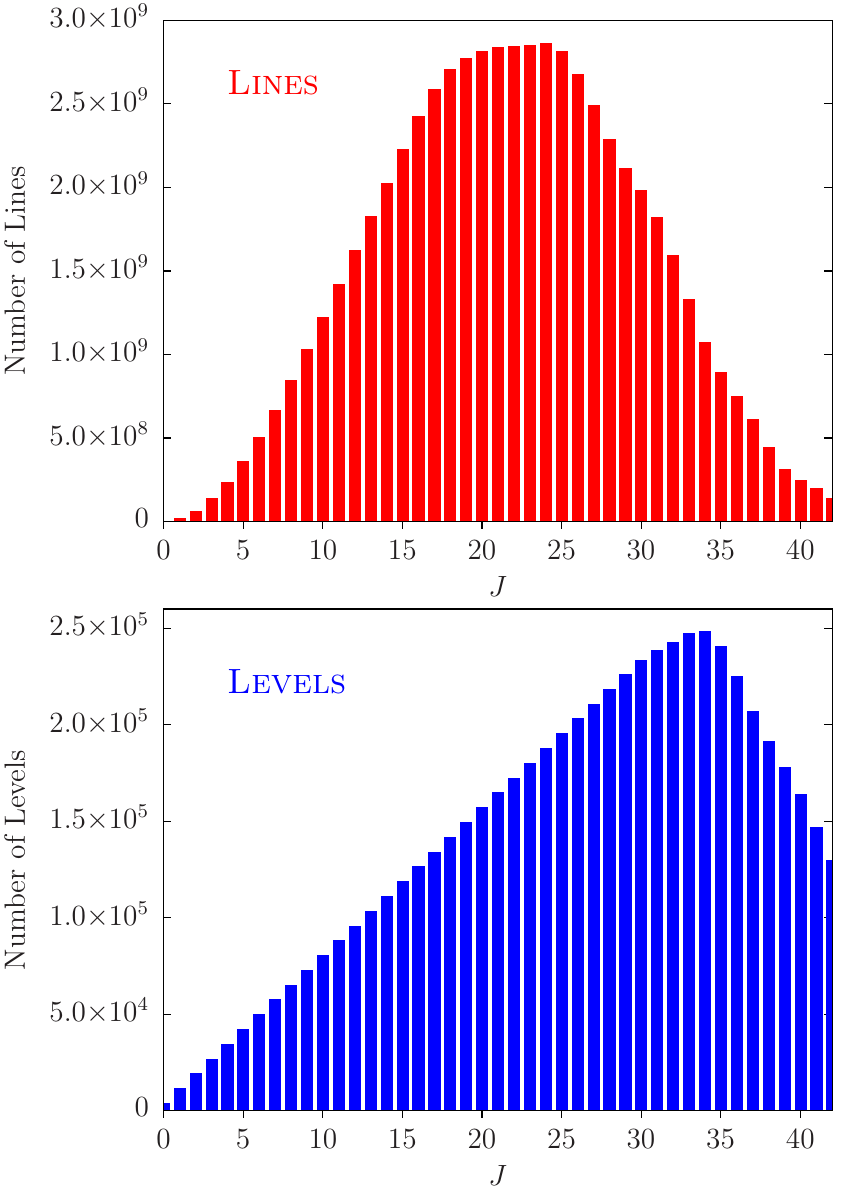}
\caption{The number of lines and energy levels in the OY2T line list for each value of the rotational quantum number $J$.}
\end{figure}

 The temperature dependence of the OY2T line list is shown in Fig.~5, where we have simulated integrated absorption cross-sections at a resolution of $1{\,}$cm$^{-1}$ using a Gaussian profile with a half width at half maximum (hwhm) of $1{\,}$cm$^{-1}$. Very weak intensities can be seen to grow several orders of magnitude stronger as the temperature increases. This smoothing of the spectrum is a result of vibrationally excited states becoming more populated and causing the rotational band envelope to broaden.

\begin{figure*}
\centering
\label{fig:temp_oy2t}
\includegraphics{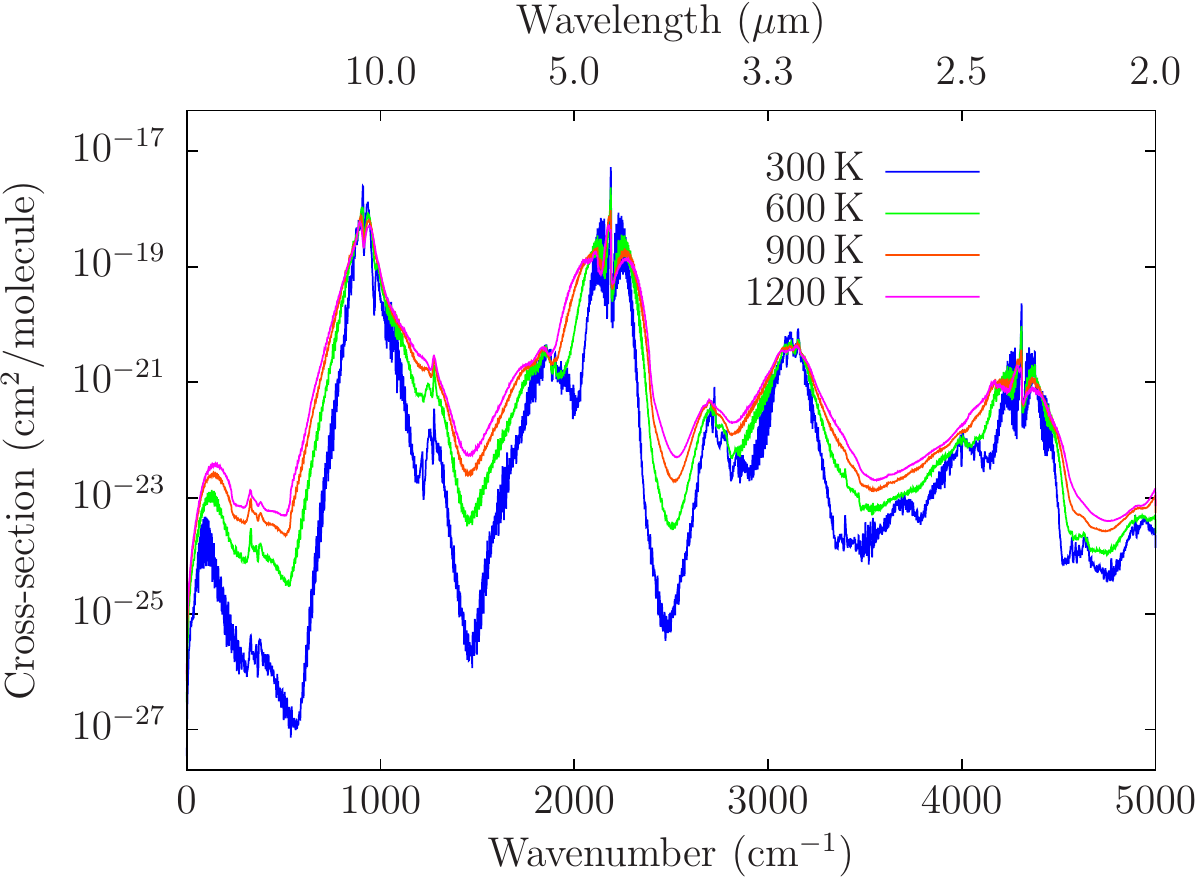}
\caption{Temperature dependence of the OY2T line list.}
\end{figure*}

 A good benchmark of the OY2T line list is to compare with the PNNL spectral library~\citep{PNNL}. In Fig.~6 and Fig.~7 we have simulated cross-sections at a resolution of $0.06{\,}$cm$^{-1}$ using a Gaussian profile with a hwhm of $0.135{\,}$cm$^{-1}$. The experimental PNNL silane spectrum was measured at a temperature of 25$\,^{\circ}$C with the dataset subsequently re-normalized to 22.84$\,^{\circ}$C ($296{\,}$K). It is of electronics grade silane gas, which is composed of $^{28}$SiH$_4$ ($92.2\%$), $^{29}$SiH$_4$ ($4.7\%$), and $^{30}$SiH$_4$ ($3.1\%$). We have therefore scaled our computed $^{28}$SiH$_4$ cross-sections by $0.922$ to ensure a reliable comparison. As expected~\citep{15OwYuYa.SiH4}, the OY2T cross-sections are marginally stronger but this is very slight. Overall the agreement is very encouraging, particularly the performance of the OY2T line list for weaker bands (see the right-hand panels of Fig.~7) as we are missing contributions from $^{29}$SiH$_4$ and $^{30}$SiH$_4$.

\begin{figure*}
\label{fig:pnnl_oy2t}
\centering
\includegraphics{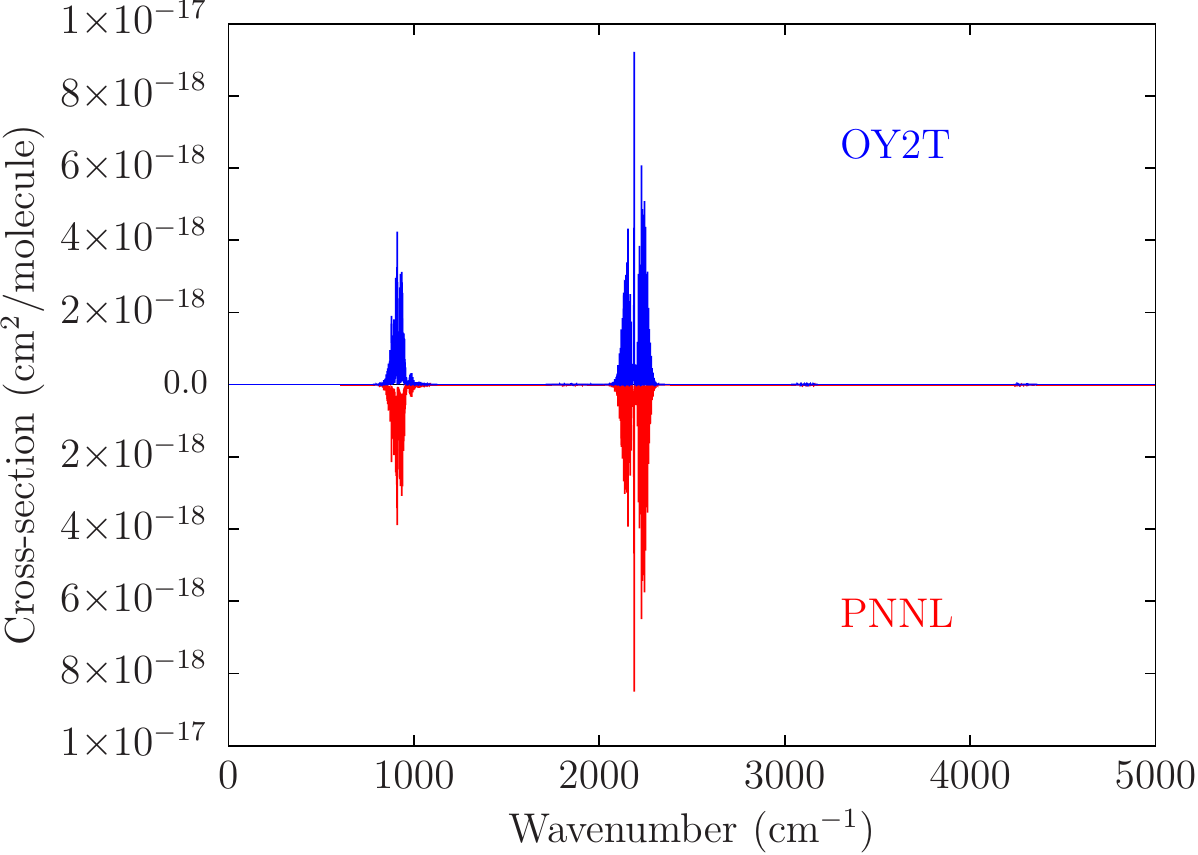}
\caption{Overview of OY2T cross-sections compared with the PNNL spectral library at $T=296\,$K. Note that the experimental PNNL spectrum~\citep{PNNL} is composed of $^{28}$SiH$_4$ ($92.2\%$), $^{29}$SiH$_4$ ($4.7\%$), and $^{30}$SiH$_4$ ($3.1\%$) (see text).}
\end{figure*}

\begin{figure*}
\label{fig:fourpanel_pnnl_oy2t}
\centering
\includegraphics{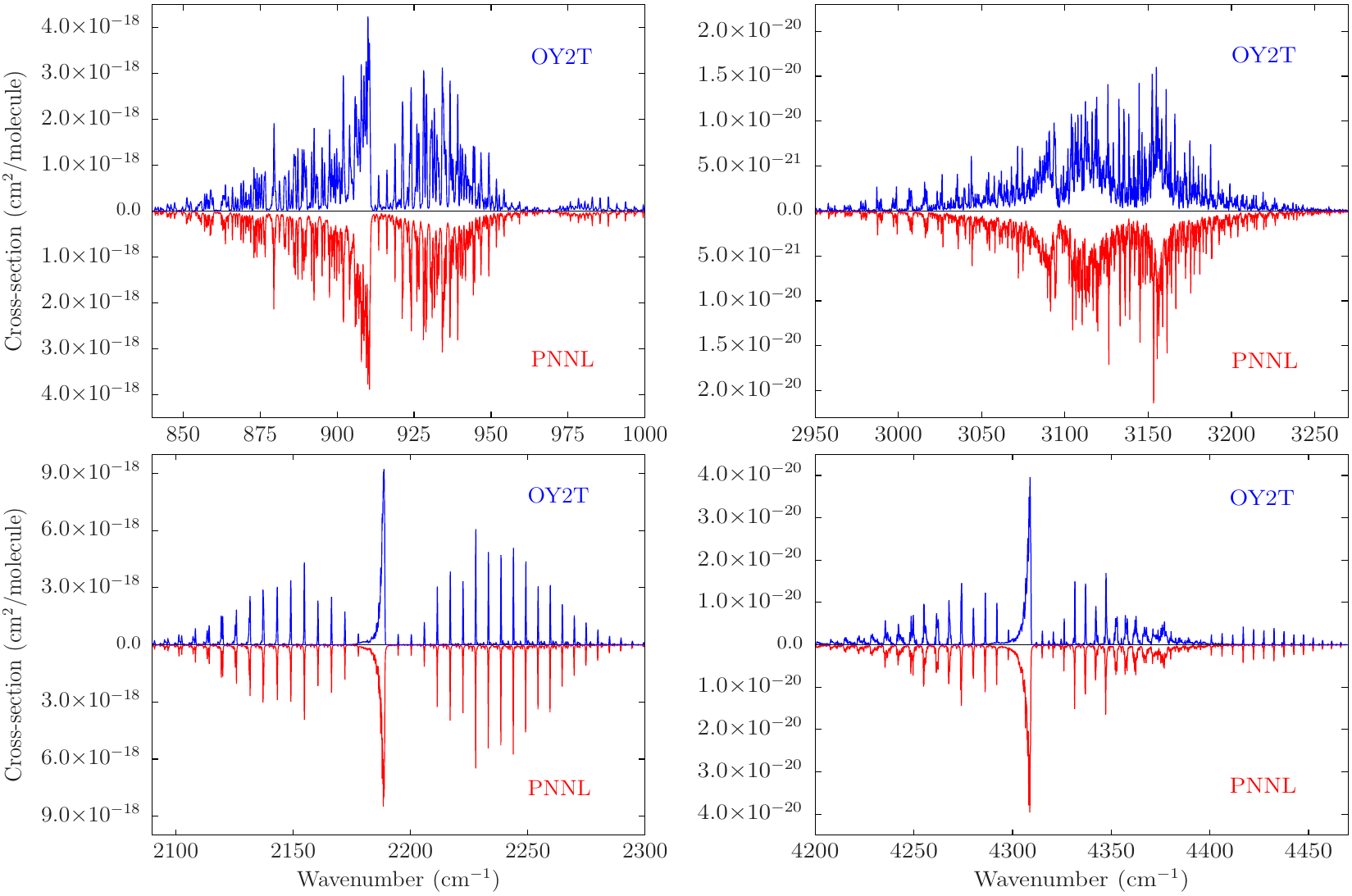}
\caption{Closer inspection of OY2T cross-sections compared with the PNNL spectral library at $T=296\,$K. Note that the experimental PNNL spectrum~\citep{PNNL} is composed of $^{28}$SiH$_4$ ($92.2\%$), $^{29}$SiH$_4$ ($4.7\%$), and $^{30}$SiH$_4$ ($3.1\%$) (see text).}
\end{figure*}

 Finally, we compare absolute line intensities of the OY2T line list with measurements of the $\nu_3$ band from \citet{15HeLoNa.SiH4}, and with the $750$--$1150\,$cm$^{-1}$ region line list from \citet{17UlGrBe.SiH4}. Intensities were computed using Eq.~\eqref{eq:abs_I} at $T=296\,$K in Fig.~8, and at $T=298\,$K in Fig.~9. As was previously found when validating the DMS~\citep{15OwYuYa.SiH4}, there is good agreement with the results of \citet{15HeLoNa.SiH4}, where transition intensities up to $J=16$ were recorded at a resolution of $0.0011{\,}$cm$^{-1}$ with an estimated experimental measurement accuracy of $10\%$. For the $^{28}$SiH$_4$ line list from \citet{17UlGrBe.SiH4} the agreement is also pleasing. Their line list contains 3512 transitions up to $J=27$ with the intensities determined from an analysis of 787 of the measured lines. Again the OY2T line intensities are slightly stronger in places but overall the band structure is well matched and weaker intensity features are accurately accounted for. It is worth noting that the OY2T line list is up to $J=42$ which would explain some of the additional spectral features in both comparisons.

\begin{figure}
\label{fig:nu3_band}
\centering
\includegraphics[width=0.47\textwidth]{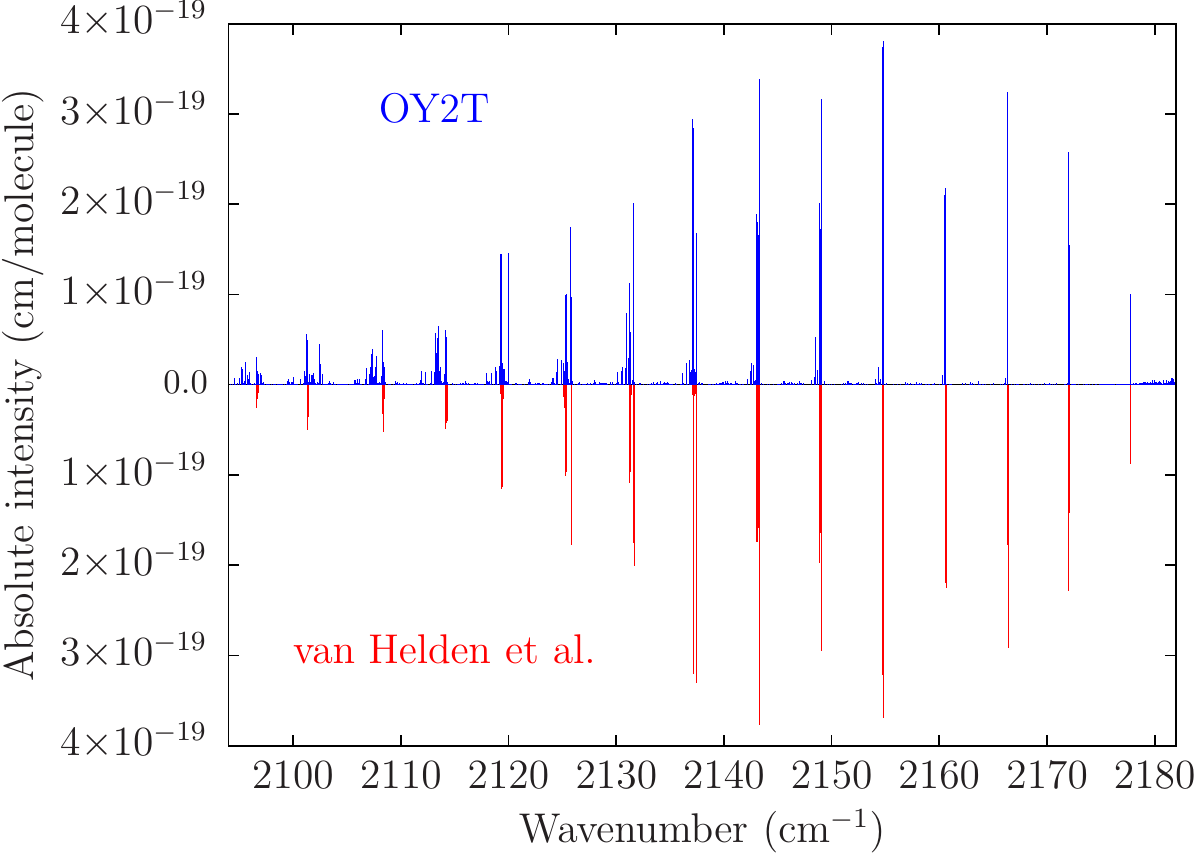}
\caption{Absolute line intensities of the OY2T line list at $T=296\,$K compared with measurements from \citet{15HeLoNa.SiH4}.}
\end{figure}

\begin{figure}
\label{fig:nu2_nu4_band}
\centering
\includegraphics[width=0.47\textwidth]{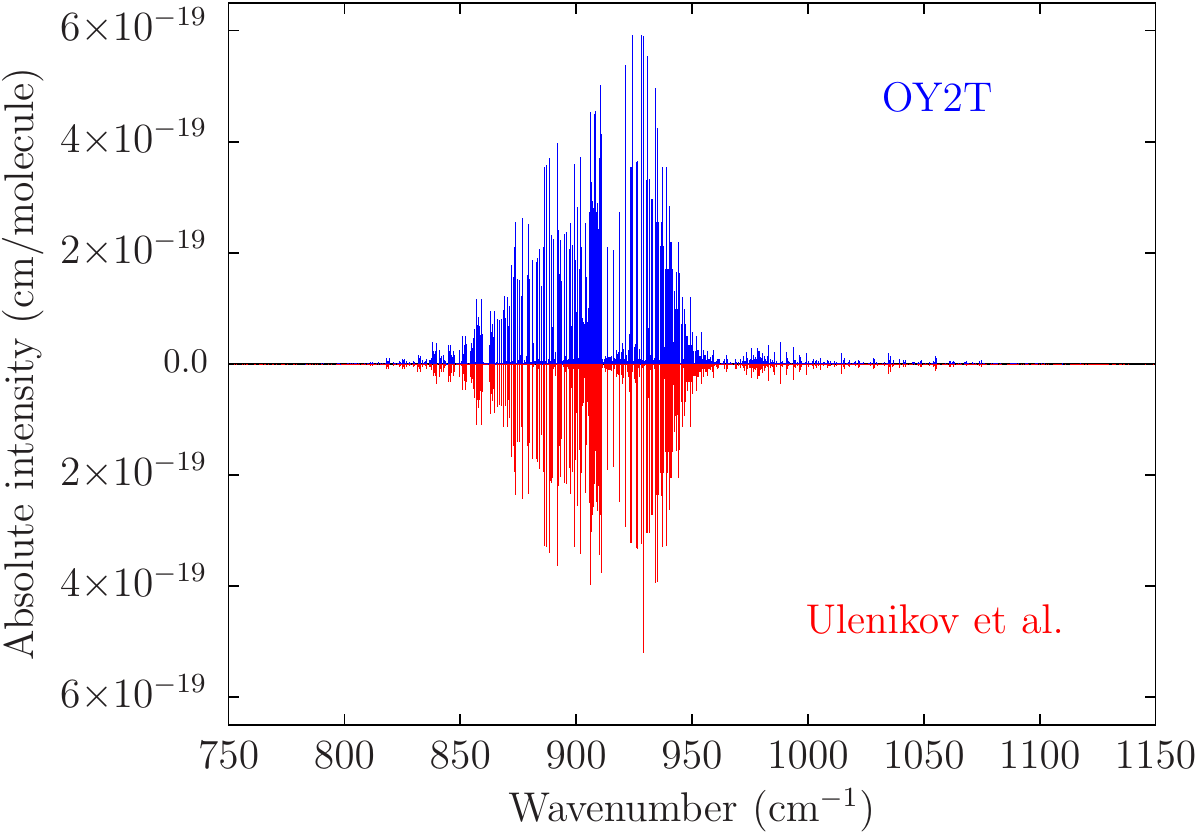}
\caption{Absolute line intensities of the OY2T line list at $T=298\,$K compared with the line list from \citet{17UlGrBe.SiH4}.}
\end{figure}

\section{Conclusion}
\label{sec:conc}

 A comprehensive rotation-vibration line list for $^{28}$SiH$_4$ has been 
presented. The OY2T line list covers the $0$--$5000\,$cm$^{-1}$ region and 
includes transitions up to $J=42$. Analysis of the temperature-dependent 
partition function suggests that the OY2T line list can be confidently used for 
temperatures below $T=1200\,$K. Applications above this temperature are likely 
to result in the loss of opacity. Comparisons with the PNNL spectral library and 
other experimental sources showed that the OY2T line list is robust and able to 
accurately reproduce the intensity features of weaker bands. The OY2T line list 
can be downloaded from the ExoMol database at \url{www.exomol.com} or the CDS 
database at \url{http://cdsarc.u-strasbg.fr}.

The line list OY2T will be useful for modelling of absorption of SiH$_4$ in 
atmospheres of exoplanets. Its quality (completeness and accuracy) should be 
sufficient, at least in principle, to detect silane in an atmosphere of a hot 
exoplanet from the transit spectroscopic observations, 
when combined with a proper atmospheric and radiative transfer models and 
providing that the abundance of SiH$_4$ is sufficiently large to be detectable. 
However for high resolution detection techniques such as the high-dispersion 
spectroscopy developed by \citet{14Snellen}, our line positions might not be 
sufficiently accurate. This technique is based on the Doppler shifts of a large 
number of spectroscopic lines of a given species, which are cross-correlated to 
the reference lab data on the line positions. The required resolution of the 
line positions should be as high, $R \geq$ 100,000. We do have a method to 
address this problem, which is to replace the theoretical energy levels in a 
synthetic line list with experimentally derived ones, usually obtained 
with the MARVEL (measured active rotational-vibrational energy levels) procedure
\citep{jt412,12FuCs.method}. The most recent example of is the hybrid line list 
for H$_3^+$ MiZATeP \citep{jt666}.  The experimental spectroscopic data 
available for SiH$_4$ should be sufficient to generate such hybrid line list for 
this molecule in the future.

 Natural extensions to the OY2T line list would be the consideration of a larger 
frequency range, an increased lower energy threshold in computations, and the 
inclusion of higher rotational excitations. These issues, although 
computationally challenging, are straightforward to address but will only be 
done if there is a demand for such work. It would be useful to derive a 
consistent set of normal mode quantum numbers  $\mathrm{v}_k$ for the OY2T line 
list. These are commonly used in high-resolution spectroscopic applications and 
could be easily incorporated by updating the \texttt{.states} file. Work in this 
direction is underway and any updates to the OY2T line list will be released on 
the ExoMol website. It may also be worthwhile to explore a more compact 
representation~\citep{ch4:2017} of the OY2T line list given the huge number of 
lines that have been generated. Certainly we are in a position to produce 
temperature-dependent cross-sections for a fixed resolution if requested.

\section*{Acknowledgments}
This work was part of ERC Advanced Investigator Project 267219. We also acknowledge support from FP7-MC-IEF project 629237, COST Action CM1405 MOLIM, and the Max Planck Computing and Data Facility (MPCDF).

\bibliographystyle{mn2e}

\section*{Supporting Information}
Additional Supporting Information may be found in the online version of this article:

\label{lastpage}

\end{document}